\documentstyle[prl,aps, graphicx]{revtex}
\begin{document}
\draft

\title{INCREASING THE BANDWIDTH OF RESONANT GRAVITATIONAL ANTENNAS: THE CASE OF EXPLORER}

\author{P. Astone$^1$, D. Babusci$^4$,  M. Bassan$^2$, P. Carelli$^3$, G.Cavallari$^9$, E. Coccia$^2$, C. Cosmelli$^5$, S.DÕAntonio$^2$, V. Fafone$^4$, A. C. Fauth$^4$,  G. Federici$^5$, G. Giordano$^4$, A.Marini$^4$  Y. Minenkov$^2$, I. Modena$^2$, G. Modestino$^4$, A. Moleti$^2$, G. V. Pallottino$^5$,  G. Pizzella$^6$, L.Quintieri$^4$,  A.Rocchi$^2$, F. Ronga$^4$, R. Terenzi$^7$, G.Torrioli$^8$, M. Visco$^7$ \\}

\address{$^1$ Istituto Nazionale di Fisica Nucleare (INFN), sezione di Roma}
\address{$^2$ Universit\`a di Roma Tor Vergata and INFN, sezione di Roma2}
\address{$^3$ Universit\`a dell' Aquila and INFN, sezione di Roma2}
\address{$^4$ Istituto Nazionale di Fisica Nucleare INFN, LNF, Frascati}
\address{$^5$ Universit\`a di Roma ÓLa SapienzaÓ and INFN, sezione di Roma}
\address{$^6$ Universit\`a di Roma ÓTor VergataÓ and INFN, LNF, Frascati}
\address{$^7$ Istituto di Fisica dello Spazio Interplanetario-CNR and INFN,sezione di Roma2}
\address{$^8$ Istituto di Fotonica e Nanotecnologie - CNR and INFN, sezione di Roma}
\address{$^9$ CERN - Gen\`eve}

\maketitle

\begin{abstract}

Resonant gravitational wave detectors with an observation bandwidth of tens of
hertz are a reality: the antenna Explorer, operated at CERN by the
ROG collaboration, has been upgraded with a new read-out. 
In this new configuration, it exhibits an unprecedented useful bandwidth: in over 55 Hz
about its frequency of operation of 919 Hz the spectral sensitivity is better than $10^{-20} Hz^{-1/2}$. We describe the detector and its sensitivity and discuss the
foreseable upgrades to even larger bandwidths.
\end{abstract}

\twocolumn 
\nopagebreak

\section{introduction}

The direct detection of gravitational waves (g.w.) using resonant detectors 
is an ongoing challenge of experimental physics.
Explorer, located at CERN since 1983, \cite{longterm,initial}  
was the first cryogenic antenna to achieve long term observations.
Other detectors operating at low (2-4 K) and ultralow (0.1-0.5 K) temperature (to reduce thermal noise, and to allow use of superconducting devices in the readout) have begun operation in the following years,
 up to the recent joint data taking of five antennas \cite{igec}.

During these years, experimenters devoted a continuous effort to improve antenna sensitivity: the first priority was to optimize the burst sensitivity,
well summarized by a unique figure: the minimum detectable energy change in the detector, that we traditionally express in terms of a pulse detection effective temperature $\Delta E_{min}=k_B T_{eff}$.

The issue of bandwidth in resonant g.w. antennas has challenged researchers since the beginning: it was soon recognized \cite{JPR84,MT84,PP84,MB84} that the useful bandwidth of such a detector is by no means limited to the 
very narrow  width of the high Q mechanical resonance. 
This 
is because the antenna has the same frequency response to both the g.w. signal and its own thermal noise (and to any mechanical excitation)
Rather, it is the amplifier noise that limits the bandwidth of the detector. In 
simplified terms,  the antenna 
is sensitive in the frequency region $\Delta f$ 
where the thermal noise is larger than the amplifier noise. As a consequence, any reduction of amplifier noise and/or increase in coupling increases the antenna bandwidth.

It is now clear that the  bandwidth of a resonant antenna
is a crucial figure of merit for sensitivity to all kind of g.w. signals 
\cite{APP97}:
in a search for periodic signals, although it does not affect the signal to noise ratio (that is set by the thermal noise only), the chance of finding a source of unknown frequency is obviously proportional to the available bandwidth.
Search for stochastic radiation is carried out by crosscorrelating the output of two detectors
\cite{stoch2}: the minimun detectable spectrum $S_h(f)$ scales with the square root of the overlapping bandwidths.

Even a search for short bursts, where 
we only care about the energy deposited in the antenna by the wave, regardless of its shape or other details, benefits of a large bandwidth: indeed, for an antenna of resonant frequency $f_a$ 
operating at temperature T,  the energy sensitivity $T_{eff}$ can be 
shown \cite {APP97} to depend on the bandwidth $\Delta f$: 
\begin{equation}
\label{Teff }
T_{eff}= \frac{f_a}{Q\Delta f} T
\end{equation}

Moreover, 
the uncertainty on the arrival time of a burst is
scales with $(\Delta f)^{-1}$: in searches with multiple detectors, an improved bandwidth allows a smaller coincidence window and therefore reduces the background of accidental coincidences. 
In the longer term, a larger bandwidth will provide more information on the incoming waveform, allowing a partial reconstruction of the signal.

The Explorer detector has been in almost continuous operation at CERN since 1991 \cite{initial}, and it has undergone over the years several overhauls that 
progressively improved both its sensitivity and its operation duty cycle. 
We report here the latest hardware upgrade, that 
has boosted the detector useful bandwidth by over an order of magnitude, up to about 55 Hz.
Explorer, like most resonant detectors, consists of a right cylinder of 5056 Al alloy of length $L=3m$, suspended in vacuum and hosted in a cryostat \cite{cryogenics} where it is cooled to about 2.6 K by a superfluid  $\ ^{4}He$ bath\cite{lambda}. A detailed description of the apparatus and its main features (including data gathering and analysis) was published in the past \cite{longterm}. Here we shall briefly recall some features relevant to the issue of bandwidth. We exploit the cylinder first longitudinal vibrational mode as "the antenna", i.e. the component that 
vibrates in response to an incoming g.w. Its effective mass (i.e. the mass of the simple harmonic oscillator that would respond with the same amplitude) is $m_a= 1135 kg$. 
To efficiently measure its amplitude of vibration, the read-out system implements two resonant  impedance matching stages: the first is  
a light-mass oscillator ($m_t =\mu m_a =0.7 kg$) \cite{paik,rapag} that is mechanically coupled to the main resonator and tuned near its resonant frequency $f_t \sim f_a=915 Hz$. The two resulting normal modes are separated by a beat frequency $\delta \geq f_a\sqrt{\mu}$ (the equal sign holds for perfect tuning, $f_a=f_t$) and contain most of the signal energy collected by the antenna and of the thermal noise of both oscillators.
The vibrational amplitude of the auxiliary oscillator modulates a d.c. bias field in a classic electrostatic transducer scheme; a superconductive transformer matches the transducer 
capacitance to the low input impedance of a high coupling, low noise d.c. SQUID.

\begin{figure}
\begin{center} 
\includegraphics[width=3in]{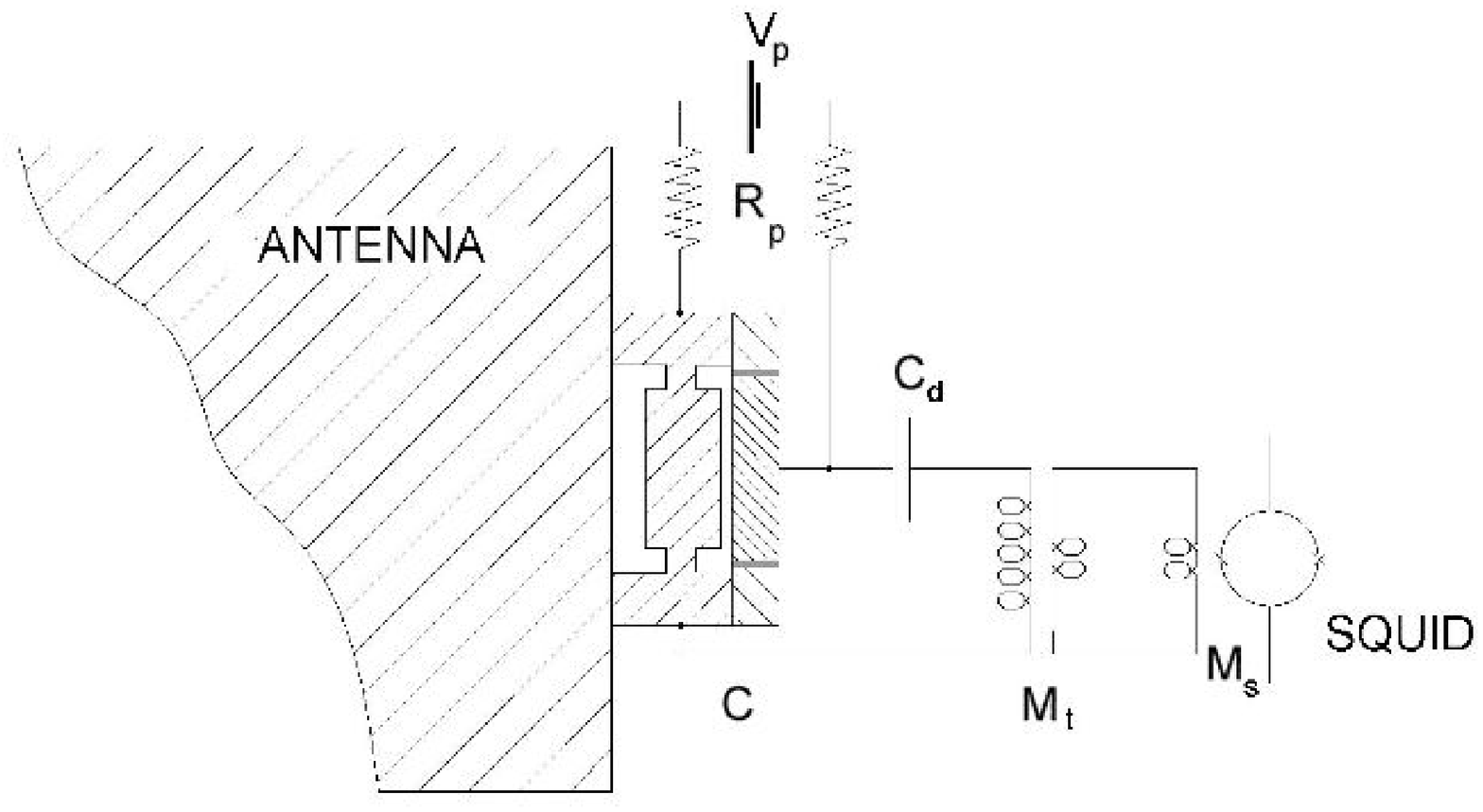}
\caption{A schematic of the Explorer detector, showing the resonant capacitive transducer and the output circuit with the superconductive transformer and the d.c. SQUID amplifier}
\label{schema}
\end{center}
\end{figure}

The second resonant matching stage is the LC circuit 
formed by the transducer capacitance and the transformer inductance (
other components, as shown in fig.\ref{schema}, and stray impedances
only account for small corrections to the resonant frequency $f_{el}$). 
A complete dynamical description of the detector and its response to both signal and noises must consider all three oscillators, with a cumbersome sixth order analysis that 
we carry out numerically. However, we have deliberately detuned the electric resonance  
( $f_{el} = 1288  Hz$, )
 to avoid that its lower Q factor degrade the mechanical ones: therefore, a two mode analysis \cite{MT84,PP84} is for most instances satisfactory; besides, as long as the  bandwidth $\Delta f$  does not exceed  the mode separation $\delta$, the two normal modes can be independently considered, and a simple, single resonance analysis \cite{APP97} can be applied. For clarity sake, we use this simplified approach to illustrate the issues, keeping in mind that it only provides intuitive guidelines, while the quantitative results shown were derived with the full numerical model.

\section{Bandwidth in resonant detectors}

We consider a single resonance antenna, with effective mass $m$ and decay time $\tau= Q/\pi f_a$. 
The mechanical vibrations are converted into an electrical signal by an apparatus that we schematize with a simple transduction coefficient $\alpha$. 
The amplifier is characterized by a wide-band output noise with spectrum $S_o$;
the resonant input noise is, for our SQUID amplifier, negligible with respect to thermal noise.
The  spectral sensitivity is easily computed \cite{APP97} by explicitely writing the ratio of the detector response to an excitation with wave amplitude $\textit{h(t)}$ and spectrum $S_h(f)$ to the spectrum of total noise, sum of thermal (resonant) plus amplifier (wideband) contributions.  By letting the $SNR(f) = 1$, the minimum detectable g.w. spectrum can be written, in terms of a normalized frequency $x=f/f_a$ :
\begin{equation}
S_h(x)=\frac{\pi k_B T}{8mQL^2 f_a^3x^4}[1+\Gamma(x^2+Q^2[1+x^2]^2)]
\label{essseacca}
\end{equation}
For the complete (three oscillators) model, the polynomial in eq. \ref{essseacca} is replaced by a higher order one.

The  coefficient
$\Gamma =\frac {S_o m \omega_a^2}{2\tau\alpha^2k_BT}$
, a dimensionless ratio of wide band to resonant noise spectra, is the crucial
parameter governing the antenna bandwidth:

\begin{equation}
\label{ deltaf}
\Delta f = \frac {f_a}{Q \sqrt{\Gamma}}
\end{equation}

This shows that a large transduction coefficient $ \alpha$ is needed in order to obtain a large bandwidth.
In the bare capacitive transducer this coefficient is simply given by the  d.c. electric field $E$ in the gap, 
but if we consider the magnetic flux coupled, through our matching circuit into the SQUID by a unit displacement we get:

\begin {equation}
\label{3.10}
\alpha_{\phi}\equiv j\omega_a C_t\frac{N_eM_s}{1- {f^2_a\over
f^2_{el}} } E     \hskip 2cm [Wb/m]
\end{equation}
 where $\omega_a C_t$ is the transducer electrical impedance, $N_e$ is the effective turn ratio of the superconducting transformer and $M_s$ is the mutual inductance between the input inductance and the SQUID.
All three parameters can be acted upon, in the design phase, to improve the antenna bandwidth.
In Explorer, we have adopted a dc SQUID with a large input coupling $M_s = 
10.6 nH$
\cite{quantumdesign} and a new resonant capacitive transducer \cite{trasd} that, due to accurate surface finishing, can be assembled with a gap as small as $10 \mu m$, resulting in a capacitance  $C_t = 12 nF$.

\section{Tuning of the detector}
It is well known \cite{paik,rapag} that the transducer resonant frequency $f_t$ is modified by the additional restoring force provided by the electrical field confined in the gap: in first approximation $\tilde{f}_t(E) = f_{t}\sqrt{1-\beta}$, where $\beta\equiv \frac{E^2C_t}{m_t\omega^2_t}$ is the figure of merit for transducer coupling. This in turn modifies the structure of the normal modes. For our system, the eigenfrequencies are given by the positive roots of the characteristic equation
\begin{eqnarray}
\label{det}
  D(f) &=& f^6-f^4\cdot[f_{el}^2+f_a^2+f_t^2(1+\mu)]\nonumber\\
&& 	+f^2\cdot[f_{el}^2f_a^2+f_{el}^2f_t^2(1+\mu)(1-\beta)+f_a^2f_t^2]\\
&&	  - f_{el}^2f_a^2f_t^2(1-\beta) =0 \nonumber
\end{eqnarray}
\begin{figure}
\begin{center} 
\includegraphics[width=3in]{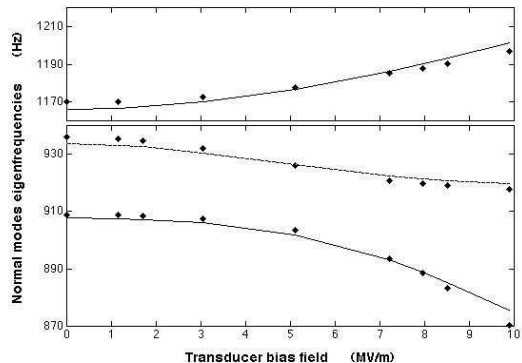}

\caption{
Tuning range of Explorer: the frequencies of the three normal modes change as the electric field in the gap (and hence 
$\beta$) is increased. The upper mode resonates well  above the others and is therefore almost purely electrical, 
 The measured values ($\diamond$) are shown 
 with the best fit to eq. \protect\ref{det}
(lines)}
\label{tuning}
\end{center}
\end{figure}

The mode frequencies of Explorer can be changed by tens of Hz,  due to the large value of $\beta$  (up to 0.05) achieved with the present transducer (see fig.\ref{tuning}). This very large tuning range makes it possible to operate the detector either with the tranducer resonant frequency well tuned  to $f_a$ or quite far from it: the two resulting lower normal modes  (the upper one is to large extent purely electrical, and bears no sensitivity to g.w.) are respectively  at a minimum separation of 24 Hz, or as wide as 44 Hz with the higher field  (9MV/m) used. As a rule of thumb, supported by detailed analysis, the cross section to g.w. of each mode is inversely proportional to the distance of its resonant frequency from $f_a = 915 Hz$. 

Tuned operation provides better mechanical matching: however, if the auxiliary oscillator has a much lower Q factor than the antenna, overall performance can be deteriorated by the resulting low Q of the normal modes.
This was the case in 2000-2002 operation, when we chose to operate in an untuned mode, by raising the  coupling field well above what needed for optimal tuning. In this way, only one normal mode retained sensitivity to g.w. and its quality factor was preserved.
In 2003 the transducer has been reconditioned, the Q of the auxiliary resonator has improved to above $2\cdot 10^6$ and tuned operation is again advantageous (see fig. \ref{tre_Sh}).

\section{Calibration and data quality}
The Explorer g.w. detector is calibrated through either a piezoelectric or a capacitive secondary transducer: the usual procedure \cite{calibration}
requires a self-calibration of these auxiliary transducers via a reflection measurement: a large pulse is sent and then observed via the same device, allowing evaluation of its coupling 
We have 
calibrated the energy pulse sensitivity of Explorer and found it to agree within $\pm 5\%$ accuracy, limited by statistical spread and by some systematics, with the values predicted by our knowledge of the model, filters and system parameters.

\begin{figure}
\begin{center}
\includegraphics[width=8cm]{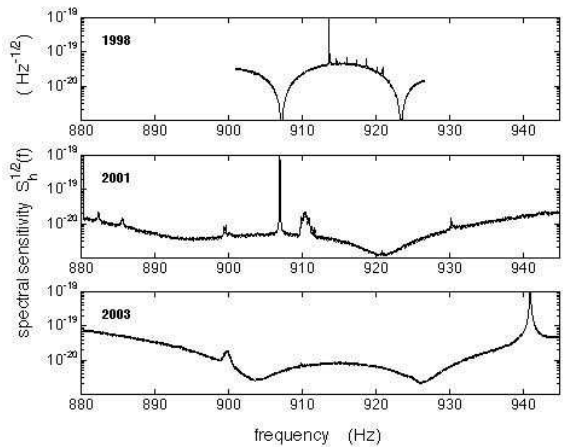}
\caption{The spectral sensitivity of the Explorer detector. Top: in 1998, before the hardware upgrade.  Middle: with increased bandwidth, in single mode operation. Bottom: a recent spectrum (Feb. 2003), with the  transducer resonator well tuned to the antenna. In all spectra, the lines rising high above mean level are due to either calibration signals or to harmonic disturbances of the power line.}
\label{tre_Sh}
\end{center}
\end{figure}

Explorer is in continuous data taking for more than 10 months a year (January is reserved by CERN to cryogenic maintenance), with an operation duty cycle exceeding 80\% of up time: the main interruption is a cryogen refill once a fortnight. Its noise energy  $T_{eff}$ has been, over the last 3 years,  between $2 \div 5$  mK, depending on the setup and on the bias field.  This corresponds to a minimum detectable wave amplitude $h_o = 3 \div 6 \cdot 10^{-19}$ for a conventional 1 ms pulse. Explorer has produced in these years data streams of unprecedented quality and sensitivity: filtering, event extraction and coincidence searches 
were described in detail \cite{coinci2}.  Its spectral sensitivity, as shown in  fig. \ref{tre_Sh},  is of the order of $\sqrt{S_h(f_a)}=2 \cdot 10^{-21}  Hz^{-1/2}$ on resonance, and it remains well below $ \cdot 10^{-20}  Hz^{-1/2}$  from 880 to 935 Hz. We have chosen to express the
observation bandwidth, rather than via the 
usual FWHM definition, as the region where the sensitivity is better than a given benchmark (here we chose $ \cdot 10^{-20}  Hz^{-1/2}$) because it carries more information about where useful signal can be collected and it better approaches the design requirement of a flat sensitivity over the widest possible span.  

\begin{figure}
\begin{center}
\includegraphics[width=8.5cm]{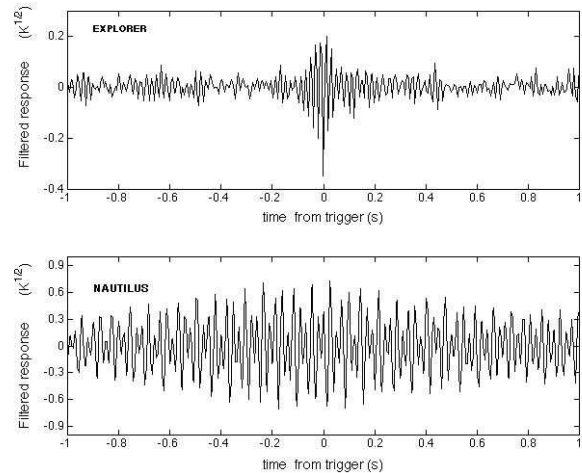}
\caption {An event triggered by a cosmic ray shower. For comparison, we show a similar event detected by the Nautilus antenna
\protect\cite{cosmici3}
where the slow beats between the two normal modes can be clearly seen: the improvement in arrival time resolution is evident}	
\label{cosmico}
\end{center}
\end{figure}

Explorer operation was suspended in Aug.2002 due to a cryogenic failure. We 
took advantage of this stop to recondition the transducer and to complete installation of a cosmic ray shower detector, consisting of layers of plastic scintillators above and below 
the cryostat, with a 
trigger logic similar to that successfully tested on Nautilus \cite{cosmici1,cosmici3}. Operation has resumed in Feb. 2003, and the first cosmic ray-triggered events have been recorded. We show in fig \ref{cosmico} the antenna response to one such event (with a multeplicity of about $3000$  particles$/ m^2$). The comparison with a similar event detected by Nautilus shows the great improvement, in terms of resolution in the arrival time, offered by a wideband readout.

\section{conclusions}
A  new readout, with a small gap capacitive transducer and a high coupling d.c. SQUID, has been successfully tested and operated on Explorer. A similar version of this readout is being  implemented on the ultracryogenic antenna Nautilus at the Frascati laboratories of INFN.

Future improvement in Explorer bandwidth and sensitivity can and will be achieved by further improving the readout: a differential capacitive transducer is presently under test, and we are confident to further reduce the wideband amplifier noise with the adoption of double stage d.c. SQUIDs that have shown \cite{AurigaSquid,harry,carelli} near quantum-limited performance. Depending on the output noise reduction that these devices will achieve, a further improvement of sensitivity, down to $10^{-4} K$, and bandwidth, up to 120 Hz, can confidently be anticipated.

\end{document}